\documentclass[prd,aps,showpacs,nofootinbib,preprint]{revtex4}
\usepackage{graphicx,color,amsmath,amsxtra}
\usepackage{epsf}
\usepackage{amssymb}
\usepackage{enumerate}
\usepackage{hhline}
\usepackage{array}
\usepackage{tabularx}
\usepackage{hangcaption}

\newcommand{\newc}{\newcommand}
\newc{\be}{\begin{equation}}
\newc{\ee}{\end{equation}}
\newc{\beq}{\begin{eqnarray}}
\newc{\eeq}{\end{eqnarray}}

\newcommand{\Eqn}[1]{&\hspace{-0.2em}#1\hspace{-0.2em}&}
\def\Vec#1{\mbox{\boldmath $#1$}}

\begin{document}

\title{Cosmological birefringence due to CPT-even Chern-Simons-like term with 
Kalb-Ramond  and scalar fields
}
\author{
Shih-Hao Ho\footnote{E-mail address: shho@mail.nctu.edu.tw},
W. F. Kao\footnote{E-mail address: gore@mail.nctu.edu.tw} 
}
\affiliation{Institute of Physics, National Chiao Tung University, Hsinchu, Taiwan 300}
\author{Kazuharu Bamba\footnote{E-mail address: bamba@phys.nthu.edu.tw}, 
C. Q. Geng\footnote{E-mail address: geng@phys.nthu.edu.tw}
}
\affiliation{Department of Physics, National Tsing Hua University, Hsinchu, Taiwan 300
}


\begin{abstract}

We study the CPT-even dimension-six Chern-Simons-like term
 by including dynamical Kalb-Ramond  and scalar fields to examine the 
cosmological birefringence. We show that the combined effect of neutrino current and Kalb-Ramond field 
could induce a sizable
rotation polarization angle in the
cosmic microwave background radiation polarization.
\end{abstract}


\maketitle

\section{Introduction}

The Lorentz and CPT invariance are foundations of particle physics. 
Testing the validity of these two invariance principles has been the hottest 
topic in the field. One of the tests is to use the cosmological birefringence~\cite{Carroll:1989vb,Feng:2006dp},
which is an additional rotation of synchrotron radiation 
from the distant radio galaxies and quasars. Since it is wavelength-independent,
  it is different from Faraday rotation.
The first indication of the cosmological birefringence was claimed by
Nodland and Ralston~\cite{Nodland:1997cc}.
Unfortunately, it has been shown that there is no statistically significant 
signal~\cite{Carroll:1997tc,Comments}.
Nevertheless, this provides a new way to search for new physics in 
cosmology. 
In recent years, there are many groups using combined data to constrain this small violation effect.
In particular, the analysis by Feng {\it et al.} gives $\Delta \alpha =-6.0 \pm 4.0 $ deg~\cite{Feng:2006dp}, while
the Wilkinson Microwave Anisotropy Probe (WMAP) group  $\Delta \alpha = -1.7 \pm 2.1$ deg 
with five year data~\cite{Komatsu:2008hk}. In addition, the Combined WMAP five year data with 
the BOOMERanG data leads to 
$\Delta \alpha = -2.6 \pm 1.9$ deg~\cite{B03,Xia:2008si}, the improved result by 
the QUaD Collaboration is 
$\Delta \alpha=0.64 \pm 0.5 \pm 0.5$ deg~\cite{Brown:2009uy}, and the combined QUaD, WMAP7, B03 and BICEP data 
indicates $\Delta \alpha=-0.04\pm0.35$ deg  ~\cite{Xia:2009ah}. 
It has pointed out that 
the Planck Surveyor~\cite{Planck} will reach a sensitivity of $\Delta \alpha$ at levels of $10^{-2}-10^{-3}$~\cite{WTNi-pol},
while a dedicated future experiment on the cosmic microwave background radiation polarization 
would reach $10^{-5}-10^{-6}$ $\Delta \alpha$-sensitivity~\cite{WTNi-pol}.

It is known that this phenomenon can be used to test the Einstein equivalence 
principle as was first pointed out by Ni~\cite{WTNi, WTNi-r}. 
Another theoretical origin of the birefringence was developed by 
Carroll {\it et al.}~\citep{Carroll:1997tc, Carroll:1989vb}. 
They modified the Maxwell Lagrangian by adding an CPT violating Chern-Simons 
term~\cite{Carroll:1989vb}, which results in numerous subsequent woks ~\cite{works}. 
In Ref.~\cite{Geng:2007va}, an CPT-even dimension-six Chern-Simons-like 
term was considered, in which
 the four-vector 
$p_{\mu}$ is related to a neutrino current \cite{Geng:2007va} and
 a Kalb-Ramond field as a auxiliary field to maintain general 
gauge invariance. 
It is clear that an observation of the cosmological birefringence may not imply CPT violation~\cite{GHN-2}
but parity violation.

In this paper, 
we 
extend the study in Ref.~\cite{Geng:2007va} by considering the dynamics of a Kalb-Ramond field 
and a scalar field. 
%
We consider the flat Friedmann-Lema\^{i}tre-Robertson-Walker (FLRW) 
space-time with the metric:
%
${ds}^2 = -{dt}^2 + a^2(t)d{\Vec{x}}^2$,
%
where $a(t)$ is the scale factor. 
%
%
We use the convention signature of the metric tensor 
$g= \mathrm{diag} (-,+,+,+)$ and 
$\epsilon^{\mu\nu\alpha\beta}=\left(1/\sqrt{g}\right) e^{\mu\nu\alpha\beta}$, 
where $e^{\mu\nu\alpha\beta}$ is the Levi-Civita tensor 
normalized by $e^{0123}=+1$. 
We also use units of $k_\mathrm{B} = c = \hbar = 1$. 

The paper is organized as follows. 
In Sec.\ II, we explain the model and derive the equations of motion. 
We explore the cosmological birefringence in Sec.\ III. 
Finally, conclusions are given in Sec.\ IV.

\section{The model}

We start with the action
\beq \label{original}
S_0 
\Eqn{=} 
\int d^4x \sqrt{g} 
\biggl[-\frac{1}{2} \epsilon \phi^2 R - \frac{1}{2}  g^{\mu\nu} \partial_{\mu} \phi \partial_{\nu} \phi -V(\phi) \nonumber  \\
& & 
- \frac{\xi_1}{6\phi^2} H_{\mu\nu\alpha} H^{\mu\nu\alpha} +\frac{\xi_2}{\phi^2} j_{\mu} \left( A_{\nu}\tilde{F}^{\mu\nu}+ \frac{1}{2}  \epsilon^{\mu\nu\alpha\beta} \partial_{\nu} B_{\alpha\beta} \right) -\frac{1}{4}  F^{\mu\nu}F_{\mu\nu}
\biggr]\,,
\eeq
where $\phi$ is the scalar field with the potential $V(\phi)$, $j_{\mu}=\bar{f}\gamma_{\mu}f\equiv (j^0,\vec{j})$ is the fermion current,
 $H_{\mu\nu\alpha} \equiv \partial_{\,[\mu}B_{\nu\alpha]}$ is 
the Kalb-Ramond field strength, $F_{\mu\nu}=\partial_{\,[\mu}A_{\nu]}$ 
and
$\tilde{F}^{\mu\nu}=\left(1/2\right) \epsilon^{\mu\nu\alpha\beta} 
F_{\alpha \beta}$ with the electromagnetic vector field $A_\mu$, and
the parameters $ \epsilon$, $\xi_1$ and $\xi_2$ are unknown constants.
It is well-known 
that Eq.~(\ref{original}) is not gauge invariant under a gauge transformation 
because of the interaction 
$\frac{\xi_2}{\phi^2} j_{\mu}\left( A_{\nu}\tilde{F}^{\mu\nu}+ \frac{1}{2} 
\epsilon^{\mu\nu\alpha\beta} \partial_{\nu} B_{\alpha\beta}\right)$. 
We note that under the gauge transformation 
$A_{\mu} \longrightarrow A_{\mu}+\partial_{\mu}\theta$, 
one obtains 
\beq 
\frac{\xi_2}{\phi^2} j_{\mu}\left( A_{\nu}\tilde{F}^{\mu\nu}+ \frac{1}{2}  \epsilon^{\mu\nu\alpha\beta} \partial_{\nu} B_{\alpha\beta}\right)  
\longrightarrow 
& &
\frac{\xi_2}{\phi^2} j_{\mu}\left( A_{\nu}\tilde{F}^{\mu\nu}+ \frac{1}{2}  
\epsilon^{\mu\nu\alpha\beta} \partial_{\nu} B_{\alpha\beta}\right) 
\nonumber \\
& & 
{}+\frac{1}{2} j_{\mu} \epsilon^{\mu\nu\alpha\beta} 
\left[ \left(\partial_{\nu} \theta\right)F_{\alpha\beta}+\partial_{\nu} \delta B_{\alpha\beta}\right]
\,.
\label{Eq2}
\eeq
The extra term in Eq. (\ref{Eq2})
from the gauge transformation should be zero, 
i.e., 
\beq \label{gauge transf}
& &
\frac{1}{2} j_{\mu} \epsilon^{\mu\nu\alpha\beta} \left[ \left(\partial_{\nu} \theta\right)F_{\alpha\beta}+\partial_{\nu} \delta B_{\alpha\beta}\right] \nonumber \\
& &
{}=
\frac{1}{2} j_{\mu} \epsilon^{\mu\nu\alpha\beta} \left[ \partial_{\nu} \left(\theta\right)F_{\alpha\beta}+\partial_{\nu} \delta B_{\alpha\beta}\right]=0\,,
\eeq
which leads to 
$\delta B_{\alpha\beta} = - \theta F_{\alpha\beta}$. 
Therefore, we have to modify the field strength tensor of $B_{\mu\nu}$ as 
\beq
\tilde{H}_{\mu\nu\alpha} \equiv H_{\mu\nu\alpha} + A_{[\mu} F_{\nu\alpha]}\,.
\eeq
As a consequence,  
the gauge invariant action becomes 
\beq \label{modified}
S_0
\Eqn{=} 
\int d^4x \sqrt{g} \biggl[-\frac{1}{2} \epsilon \phi^2 R - \frac{1}{2}  g^{\mu\nu} \partial_{\mu} \phi \partial_{\nu} \phi -V(\phi) \nonumber  \\
& & 
{}- \frac{\xi_1}{6\phi^2} \tilde{H}_{\mu\nu\alpha} \tilde{H}^{\mu\nu\alpha} +\frac{\xi_2}{\phi^2} j_{\mu}\left( A_{\nu}\tilde{F}^{\mu\nu}+ \frac{1}{2}  \epsilon^{\mu\nu\alpha\beta} \partial_{\nu} B_{\alpha\beta}\right) -\frac{1}{4}  
F^{\mu\nu}F_{\mu\nu}\biggr]\,. 
\eeq
By varying the action with respect to $\phi$, $g_{\mu\nu}$, $B_{\mu\nu}$ and $A_{\mu}$, 
we can have a set of equations of motion as follows: 
\beq \label{vary phi}
\epsilon \phi R = D_{\mu} \partial^{\mu}\phi - \frac{\partial V}{\partial \phi} + \frac{\xi_1}{3\phi^3} \tilde{H}^2 -2 \frac{\xi_2}{\phi^3} j_{\mu}\left( A_{\nu}\tilde{F}^{\mu\nu}+ \frac{1}{2}  \epsilon^{\mu\nu\alpha\beta} \partial_{\nu} B_{\alpha\beta}\right)\,,
\eeq
\beq \label{vary g}
\epsilon \phi^2 G_{\mu\nu} 
\Eqn{=}
\left[\frac{1}{2} \left(\partial_{\alpha}\phi\right)^2+V(\phi)\right]g_{\mu\nu} -\partial_{\mu}\phi \partial_{\nu} \phi + \frac{\xi_1}{6\phi^2} \tilde{H}^2 g_{\mu\nu} +\left(\frac{1}{4} F^2 g_{\mu\nu} - F_{\mu\alpha}F_{\nu}^{\alpha} \right) 
\nonumber \\
& &
{}+ \epsilon(D_{\nu}D_{\mu}\phi^2-D^{\sigma}D_{\sigma}\phi^2 g_{\mu\nu}) 
-\frac{1}{\phi^2}
\tilde{H}_{\mu\alpha\beta}\tilde{H}_{\nu}^{\alpha\beta}\,, 
\eeq
\beq \label{vary B}
D_{\mu}\left(\frac{\xi_1}{\phi^2}\tilde{H}^{\mu\nu\alpha}+\frac{\xi_2}{2\phi^2}\epsilon^{\mu\nu\alpha\beta}j_{\beta}\right)=0\,,
\eeq
\beq \label{vary A}
D_{\nu}F^{\nu\mu}-D_{\nu}\left(\frac{2\xi_1}{\phi^2}\tilde{H}^{\nu\alpha\mu}A_{\alpha}+\frac{\xi_2}{\phi^2}\epsilon^{\beta\alpha\nu\mu}j_{\beta}A_{\alpha}
\right)=\frac{\xi_1}{\phi^2}\tilde{H}^{\mu\nu\alpha}F_{\nu\alpha}-\frac{\xi_2}{\phi^2}j_{\nu}\tilde{F}^{\nu\mu}\,.
\eeq
Since $\tilde{H}^{\mu\nu\alpha}$ is a totally antisymmetric tensor, 
we can write 
$\tilde{H}^{\mu\nu\alpha}=\epsilon^{\mu\nu\alpha\beta} T_{\beta}$, 
where $T_{\beta}$ is a vector with mass dimension three. 
Thus, Eq.~(\ref{vary B}) 
is rewritten to 
\beq \label{torsion curl}
%
%
\epsilon^{\mu\nu\alpha\beta} \partial_{\mu}\left(\frac{\xi_1}{\phi^2}T_{\beta}+\frac{\xi_2}{2\phi^2} j_{\beta}\right)=0\,. 
\label{eq:A-10}
\eeq
%
Focusing on the space time manifold with first trivial homology group, any closed one-form is an exact one-form. Therefore, 
from Eq.~(\ref{eq:A-10}), we can
express the torsion field as 
\beq \label{torsion}
\frac{1}{\phi^{2}}
\left(\xi_1 T_{\beta}+\frac{\xi_2}{2} j_{\beta}\right)=
\partial_{\beta} 
\Phi\,, 
\eeq
where 
$\Phi$ is a dimensionless pseudo-scalar. With the help of 
Eq.~(\ref{torsion}), we can further simplify the equations of motion to be 
\beq 
\epsilon \phi R
\Eqn{=} 
D_{\mu} \partial^{\mu} \phi -\frac{\partial V}{\partial \phi}-\frac{2\phi}{3\xi_1}\left(\partial_{\mu} \Phi\right)^2+\frac{\xi_2^2}{2\xi_1 \phi^3}\left(j_{\mu}\right)^2\,, 
\label{eom phi} \\
%
\label{eom g}
\epsilon \phi^2 G_{\mu\nu}
\Eqn{=} 
\left[\frac{1}{2}\left(\partial_{\alpha} \phi\right)^2+V(\phi)\right]g_{\mu\nu}-\partial_{\mu}\phi \partial_{\nu}\phi +\epsilon(D_{\nu}D_{\mu}\phi^2-D^{\sigma}D_{\sigma}\phi^2 g_{\mu\nu}) \nonumber \\
& &
{}+
\frac{1}{\xi_1 \phi^2} \left[\phi^4\left(\partial_{\alpha}\Phi\right)^2-\xi_2 
\phi^2 j_{\alpha} \partial^{\alpha} \Phi +\frac{\xi_2^2}{4}\left( j_{\alpha} 
\right)^2 
\right] g_{\mu\nu} \nonumber \\
& &
{}+ 
\left(\frac{1}{4}F^2 g_{\mu\nu}-F_{\mu\alpha}F_{\nu}^{\alpha}\right)-2\frac{\xi_1}{\phi^2}
\left(\frac{\phi^2}{\xi_1}\partial_{\mu}\Phi - \frac{\xi_2}{2\xi_1}j_{\mu}\right) \left(\frac{\phi^2}{\xi_1}\partial_{\nu}\Phi - \frac{\xi_2}{2\xi_1}j_{\nu}\right) \,,  
\label{eom g} \\
%
D_{\mu}F^{\mu\nu}
\Eqn{=} 
-4 \left(\partial_{\mu} 
\Phi\right) \tilde{F}^{\mu\nu}\,.
\label{eom A}
\eeq
%

\section{Cosmological birefringence}

Now, we consider the simplest $\phi^4$ potential for the scalar with both $V_0$ and 
$\lambda$ larger than zero 
\beq \label{potential}
V(\phi)
=\lambda\left(\phi^2-\phi_0^2\right)^2+V_0 \ .
\eeq

Defining $\phi_0^2=m^2/ \left(2\lambda\right)$ and taking all the $\phi$ field 
in the equations to be  $\phi_0$,  equations of motion become 
\beq  
\epsilon \phi_0 R
\Eqn{=}
-2\frac{\phi_0}{\xi_1}\left(\partial_{\mu} \Phi\right)^2+\frac{\xi_2^2}{2\xi_1 \phi_0^3}\left(j_{\mu}\right)^2\,, 
\label{eom phi 0} \\
%
\epsilon \phi_0^2 G_{\mu\nu}
\Eqn{=} 
V_0g_{\mu\nu}+\left(\frac{1}{4}F^2 g_{\mu\nu}-F_{\mu\alpha}F_{\nu}^{\alpha}\right) \nonumber \\
& &
{}+
\frac{1}{\xi_1 \phi_0^2}\left[\phi_0^4\left(\partial_{\alpha}\Phi\right)^2-\xi_2 \phi_0^2 j_{\alpha} \partial^{\alpha} \Phi +\frac{\xi_2^2}{4}\left( j_{\alpha}\right)^2\right] g_{\mu\nu} \nonumber \\
& &
{}-
2\frac{\xi_1}{\phi_0^2}
\left(\frac{\phi_0^2}{\xi_1}\partial_{\mu}\Phi - \frac{\xi_2}{2\xi_1}j_{\mu}\right) \left(\frac{\phi_0^2}{\xi_1}\partial_{\nu}\Phi - \frac{\xi_2}{2\xi_1}j_{\nu}\right) \,.
\label{eom g 0}
\eeq
%
%
Taking the trace of Eq.~(\ref{eom g 0}), we have
\beq \label{contract}
 -\epsilon \phi_0^2 R 
\Eqn{=} 
4V_0+2\frac{\phi_0^2}{\xi_1}\left(\partial_{\mu} \Phi\right)^2 -2\frac{\xi_2}{\xi_1} j_{\mu} \partial^{\mu} \Phi +\frac{\xi_2^2}{2\xi_1 \phi_0^2}\left(j_{\mu}\right)^2\,.
\eeq
By combining Eqs.~(\ref{eom phi 0}) and (\ref{contract}), we obtain 
\beq \label{final}
&&4V_0-2\frac{\xi_2}{\xi_1}j_{\mu}(\partial^{\mu}\Phi)+\frac{\xi_2^2}{\xi_1\phi_0^2}(j_{\mu})^2=0\,.
\eeq
%


In the FLRW Universe, it is reasonable to 
assume a homogeneous and isotropic fermion current and torsion field~\cite{Geng:2007va}, i.e., 
$j_{\mu}=( j_0 (t), 
\vec{0})$ and $T_{\mu}=( T_0 (t), \vec{0})$. From Eq.~(\ref{final}), 
we have the evolution equation for the dimensionless psedo-scalar 
$\Phi$: 
\beq \label{eom lambda}
4V_0+2\frac{\xi_2}{\xi_1}j_0(\partial_0\Phi)-\frac{\xi_2^2}{\xi_1\phi_0^2}(j_0)^2=0\,.\eeq
The solution of Eq.~(\ref{eom lambda}) can be easily derived as 
\beq \label{sol lambda}
\partial_0 \Phi =-2\frac{\xi_1V_0}{\xi_2 j_0}+\frac{\xi_2 }{2\phi_0^2}j_0\,. 
\eeq
%
Similar to the calculation in Ref.~\cite{Geng:2007va}, the change in the position 
angle of the polarization plane $\Delta \alpha$ at the redshift  
$z \equiv 1/a -1$ is given by 
\beq 
\Delta \alpha 
\Eqn{=} 
2\int \left(\partial_0 \Phi\right) \frac{d t}{a(t)} = 
2\int_0^{1100} \left(-2 \frac{\xi_1V_0}{\xi_2 j_0}+\frac{\xi_2 }{2\phi_0^2}j_0\right) \frac{d z}{H_0 \left(1+z\right)^{3/2}} 
\label{angle1}
\eeq
%
where
$H_0=2.1 \times 10 ^{-42}h$\,GeV is the Hubble constant with 
$h\simeq 0.7$ at the present and we have assumed our Universe is flat and 
matter-dominated.
%
To estimate $\Delta \alpha $ in Eq. (\ref{angle1}), we take
the zero component of the 
fermion current $j_0$ to be the (lightest) neutrino asymmetry, say, 
the electron neutrino in our universe,
\beq \label{neutrino}
j_0=\Delta n_{\nu_e}=\frac{1}{12\zeta(3)}\left(\frac{T_{\nu_e}}{T_{\gamma}}\right)^3 \pi^2 \xi_{\nu_e} n_{\gamma}=\frac{2}{33} \xi_{\nu_e} T_{\gamma _{0}}^3 
\left(1+z\right)^3\,, 
\eeq
where $T_{\gamma_0}$  is the CMB temperature at the present, $\xi_{\nu_e}$ is the degeneracy parameter for 
the electron neutrino and 
$\left(T_{\nu_e}/T_{\gamma}\right)^3=4/11$ is assumed. 
In Ref.~\cite{Serpico:2005bc} the bound on the degeneracy parameter is 
$-0.046 < \xi_{\nu_e} < 0.072$ for a $2\sigma$ range of the baryon asymmetry. 

%
%

Inserting Eq.(\ref{neutrino}) in to Eq.(\ref{angle1}), we have
\beq \label{angle2}
\Delta \alpha = 
2 f(z){\large |}^{1100}_0
\eeq
where $f(z)$ is given by
\beq \label{anglef}
f(z) = 
\left(\frac{66\xi_1 V_0}{7 \xi_2\xi_{\nu_e} T^3_{\gamma_0}H_0}  \right) (1+z)^{-7/2}  + \left( \frac{2 \xi_2 \xi_{\nu_e} T^3_{\gamma_0} }{165\phi_0^2H_0} \right) (1+z)^{5/2} .
\eeq
Therefore, there is a bound of the function $|f(z)|$
\beq \label{inequality}
|f(z)|  \ge 2 \left[ { 4 \xi_1 V_0 \over 35 \phi_0^2  H_0^2 (1+z)} \right  ]^{1/2},
\eeq
which can be thought of as a bound on the contribution of the effective cosmological constant $V_0$.
As an illustration,
 for example, by taking $\phi_0=M_{pl}=\sqrt{1/8\pi G}$ the reduced Planck mass(taking $\epsilon=1$ hereafter), $V_0\sim 10^{-85}$\, (GeV)$^4$, $\xi_1=1$, $\xi_2=1$ and
$\xi_{\nu_{e}}\sim 10^{-3}$, we get $\Delta \alpha
\sim -9.7 \times 10^{-2}$, 
which could explain the results in 
Refs.~\cite{Feng:2006dp, B03, Komatsu:2008hk, Xia:2008si, Brown:2009uy}.
\section{Conclusions}

In the present paper, we have studied 
the CPT-even dimension-six Chern-Simons-like term
in Ref.~\cite{Geng:2007va} by including dynamical torsion  and scalar fields to explain the 
cosmological birefringence effect. The combined effect of the Kalb-Ramond field and  neutrino current  induces a sizable
rotation polarization angle in the CMB data provided that there is a non-zero
neutrino number asymmetry. 

It is interesting to note that the effect induced by the Kalb-Ramond field is  the inverse of the one due to the neutrino current, as shown in Eq.~(\ref{anglef}). In contrast to the model in Ref.~\cite{Geng:2007va}, in which  a similar dimension-six interaction with an undetermined effective coupling constant was examined, we consider, however, the dynamical scalar field as the 
coupling constants of the
Ricci scalar, Kalb-Ramond  field and interaction terms. Namely, the effective coupling constant $\phi_0^2$ is related to $M_{pl}$ in the Einstein-Hilbert action. Because of this limitation, the contribution to the angle $\Delta \alpha$ is highly suppressed to $O(10^{-32})$, and the corresponding $V_0$ has to be around
$10^{-85}$ (GeV)$^4$ to match the current observational constraint.

Finally, we remark that there should be other 
interesting cosmological phenomenology in this model \cite{Kao:1992xz}, which will be studied elsewhere.

\section*{Acknowledgments}
The work by S.H.H. and W.F.K. is supported in part by
the National Science Council of R.O.C. under
Grant number: NSC-98-2112-M-009-002-MY3 and that 
 by K.B. and C.Q.G. is supported in part by 
the National Science Council of R.O.C. under
Grant numbers: NSC-95-2112-M-007-059-MY3 and
NSC-98-2112-M-007-008-MY3
and 
National Tsing Hua University under the Boost Program and Grant \#: 
99N2539E1.




\end{document}